\documentclass[aps,pra,twocolumn,groupedaddress,showpacs]{revtex4}

\usepackage{graphicx}
\usepackage{amsmath}

\begin{document}


\title{Three-dimensional dynamics of vortex lattice formation in Bose-Einstein condensates}


\author{Kenichi Kasamatsu$^1$}
\author{Masahiko Machida$^2$}
\author{Narimasa Sasa$^2$}
\author{Makoto Tsubota$^3$}
\affiliation{$^1$Department of General Education, 
Ishikawa National College of Technology, Tsubata, Ishikawa 929-0392, Japan\\
$^2$CCSE, Japan Atomic Energy Research Institute, 6-9-3 
Higashi-Ueno, Taito-ku, Tokyo 110-0015 Japan \\
$^3$Department of Physics,
Osaka City University, Sumiyoshi-Ku, Osaka 558-8585, Japan}

\date{\today}

\begin{abstract}
We numerically study the dynamics of vortex lattice formation in a rotating cigar-shaped Bose-Einstein condensate. The study is a three-dimensional simulation of the Gross-Pitaevskii equation with a phenomenological dissipation term. The simulations reveal previously unknown dynamical features of the vortex nucleation process, in which the condensate undergoes a strongly turbulent stage and the penetrating vortex lines vibrate rapidly. The vibrations arise from spontaneous excitation of Kelvin waves on the proto-vortices during a surface wave instability, caused by an inhomogeneity of the condensate density along the elongated axial direction. 
\end{abstract}


\maketitle
Atomic-gas Bose-Einstein condensates (BECs) provide a versatile testing ground for superfluidity, particularly in systems with an externally driven rotation. Several experimental groups have observed the formation of a lattice of quantized vortices in such a rotating BEC \cite{Madison,Abo,Haljan,Hodby}. The experiment that largely motivates the present study is a {\it direct} observation of nonlinear dynamical phenomena such as vortex nucleation and lattice formation of a rotating BEC in real time \cite{Madison2}. There are some theoretical attempts to understand the dynamical properties of vortex lattice formation in this system \cite{Feder,Tsubota,Penckwitt,Kasamatsu,Lundh,Lobo,Sasa}. Among these studies, the experimental observations are well reproduced by two-dimensional (2D) numerical simulations of the time-dependent Gross-Pitaevskii equation (GPE) with a {\it phenomenological dissipation term} \cite{Tsubota,Penckwitt,Kasamatsu,Choi}. 

This paper presents the full three-dimensional (3D) dynamics of vortex lattice formation in a rotating BEC through the numerical simulations of the time-dependent GPE with a dissipation term. Except for a few preliminary works \cite{Feder,Lobo,Sasa} and studies of the static configurations of vortices \cite{Ripoll,Feder2}, this problem has been analyzed using only 2D simulations. In contrast to 2D simulations, a 3D simulation gives significant improvement for describing real 3D systems. For the present study, these improvements are (i) the 3D simulation includes the axial ($z$) component, which increases the number of degrees of freedom of the fluctuation modes, and (ii) the 3D simulation can treat observed physical phenomena associated with ``vortex lines", such as bending \cite{Ripoll}, Kelvin waves (helical displacements of the vortex core) \cite{Bretin}, and reconnection \cite{Koplik}. These features make the dynamical process of vortex lattice formation richer than that of the 2D case. Recently, Lobo {\it et al.} studied the 3D dynamics of rotating atomic-gas BECs using classical field theory in which the dynamics is described by the energy-conserving time-dependent GPE and both the condensate wave function and its fluctuation modes are approximated by a single c-number field \cite{Lobo}. The dissipative process and the 3D vortex dynamics of a rotating BEC in a spherically-symmetric trap was discussed in Ref. \cite{Sasa}. In our present study, we take account of the phenomenological dissipation so as to reproduce the entire vortex formation process that was observed in the experiments \cite{Madison2,tyuu}.

In our previous numerical study \cite{Tsubota,Kasamatsu}, we described four dynamical stages of vortex lattice formation: (i) The condensate undergoes a large-amplitude quadrupole oscillation that is damped because of the dissipation. (ii) After a few hundred milliseconds, the boundary surface of the condensate becomes unstable, generating surface ripples that propagate along the surface. (iii) The surface ripples develop into the vortex cores, which enter the condensate. (iv) The vortices form an ordered lattice and the global shape of the condensate recovers the axisymmetry. Particularly, in stages (ii) and (iii), the surface ripples are connected with proto-vortices, named ``ghost vortices" \cite{Tsubota}, that appear in the low density region outside the condensate. These processes are consistent with the observation in Ref. \cite{Madison2}, but the dynamics of the nucleating vortex lines themselves has not been reported yet. However, the above simulation was 2D, thus differing from the actual 3D system. The present 3D simulations add some new qualitative aspects to the vortex dynamics, especially when the condensate undergoes a surface wave instability. 

Here we assume a cigar-shaped condensate because this is the case that was studied experimentally by Madison {\it et al.} \cite{Madison,Madison2}. In this study, we find that the transitional process from the nonvortex state to the vortex state is governed by the strongly turbulent dynamics of the condensate wave function. This turbulence arises from {\it spontaneous Kelvin wave excitation of the ghost vortex lines} before the lines penetrate into the condensate. The Kelvin waves are excited because of the density inhomogeneity along the $z$-axis. Therefore, vortex lines also vibrate strongly immediately after they penetrate into the condensate. This significantly surpresses the visibility of the vortex cores in the 2D column density $\rho(x,y) = \int dz \rho(x,y,z)$ which is often used to visualize the vortices experimentally.

A BEC trapped in an external potential $V({\bf r})$ is described by a condensate wave function $\Psi({\bf r},t)$ obeying the GPE. In a frame rotating with a frequency $\Omega$ around the $z$-axis, the generalized GPE with a phenomenological dissipation constant $\gamma$ reads \cite{Tsubota,Penckwitt,Kasamatsu,Choi}
\begin{equation}
(i-\gamma) \frac{\partial \Psi}{\partial t} = \biggl( - \nabla^{2} + \tilde{V} - \mu + u |\Psi|^{2} - \Omega \hat{L}_{z} \biggr) \Psi, \label{gpedimless}
\end{equation} 
where $\mu$ is the chemical potential, $\hat{L}_{z}=-i (x \partial_{y} - y \partial_{x})$ the angular momentum operator and the wave function is normalized to unity: $\int d{\bf r} |\Psi|^{2} =1$. The units of energy, length and time are given by the corresponding scales of the radial harmonic potential as $\hbar \omega_{\perp}$, $b_{\perp}=\sqrt{\hbar/2m\omega_{\perp}}$ and $\omega_{\perp}^{-1}$, respectively, where $m$ is the atomic mass and $\omega_{\perp}$ the radial trap frequency. The trapping potential is $\tilde{V}=[(1+\epsilon) x^{2} + (1-\epsilon) y^{2}]/4 + \lambda^{2} z^{2}/4$ with the aspect ratio $\lambda=\omega_{z}/\omega_{\perp}$ and a small anisotropy $\epsilon$; this form describes approximately the rotating potential used in the Madison {\it et al.} experiments \cite{Madison,Madison2}. The strength of the mean-field interaction between atoms is characterized by $u$, which is proportional to the $s$-wave scattering length $a$ as $u=8 \pi N a /b_{\perp}$, where $N$ is the total particle number. 

We will compare the present results with 2D simulation results to clarify their differences. To reduce the dimension from 3D to 2D, we assume translation symmetry along the $z$ direction. Since we will be comparing to experiments on a condensate in an elongated, cigar-shaped trap with $\lambda \simeq 10^{-1} \sim 10^{-2}$ \cite{Madison,Madison2}, we assume that the $z$-axis is along the symmetry axis of the experiments. We obtain the 2D formulation from the 3D approach by separating the degrees of freedom of the wave function as $\Psi({\bf r},t)=\psi(x,y,t)\phi(z)$ \cite{Tsubota,Kasamatsu}. As a result, we obtain the 2D version of Eq. (\ref{gpedimless}) with the alternating coupling constant $u \rightarrow u_{\rm 2D}=8 \pi a N \int dz |\phi(z)|^{4}/(\int dz |\phi(z)|^{2})^{2} \simeq 8 \pi a N h_{z}$, where $h_{z}$ is a height of the cylinder. 

Because a vortex lattice state corresponds to a local minimum of the total energy functional, a phenomenological dissipation constant $\gamma$ has been introduced in Eq. (\ref{gpedimless}) by replacing $i\partial/\partial t$ with $(i-\gamma)\partial/\partial t$ \cite{Tsubota,Kasamatsu,Choi}. We assume $\gamma =0.03$, which was obtained by Choi {\it et al.} \cite{Choi} by fitting a numerical simulation of the generalized GPE with the experimental data on collective excitations in Ref. \cite{Mewes}. Compared with the other terms in the GPE, this dissipation constant is so small that any small variation of $\gamma$ could not change the dynamical process qualitatively. Because the time development of the GPE with $\gamma$ does not conserve the norm of the wave function, we adjust the chemical potential $\mu$ at each time step to ensure normalization. This is done by calculating the correction $\Delta \mu = (\Delta t)^{-1} \ln [ \int d {\bf r} |\Psi(t)|^{2} /\int d {\bf r} |\Psi(t+\Delta t)|^{2} ]$ \cite{Jackson}. Since the dissipation term can be derived from the formulation for a finite-temperature BEC developed by Zaremba {\it et al.} \cite{Zaremba} (under some approximations \cite{Kasamatsu}), the dissipation should be related to the presence of a thermal component. Gardiner {\it et al.} obtained a similar but different GPE with the dissipative term using quantum kinetic theory \cite{Gardiner}, applying it to their simulation of vortex lattice formation \cite{Penckwitt}. For a fixed chemical potential, the dynamics in the present study is exactly coincident with that in another model in Ref. \cite{Penckwitt} provided that the atomic cloud and the trap are rotated at the same angular frequency and the gauge transformation $\Psi \rightarrow \Psi e^{-i \mu t / \hbar}$ is made. 

The numerical scheme to solve Eq. (\ref{gpedimless}) is a Crank-Nicholson method with the spatial mesh $\Delta_{x,y,z} = 0.25$ and the time step $\Delta_{t} = 5 \times 10^{-4}$. We use the ground state solution of Eq. (\ref{gpedimless}) with a nonrotating axisymmetric trap ($\epsilon=0$) as the initial state of the simulations. The conditions of the Madison {\it et al.} experiments on $^{87}$Rb atoms \cite{Madison2} are the following: $N=3 \times 10^{5}$, $\omega_{\perp} = 108.56 \times 2 \pi $ Hz, $\omega_{z} = 11.8 \times 2 \pi$ Hz ($\lambda=9.2^{-1}$), and $a=5.29$ nm, which yields the length scale $b_{\rm ho}=0.728$ $\mu$m, the time scale $\omega_{\perp}^{-1}=$1.47 msec, and the dimensionless coupling constant $u=54760$. In the 2D calculation, we obtain the 2D coupling constant $u_{\rm 2D}=440$, where the height of the cylinder $h_{z}$ (along the $z$-axis) is assumed to be the Thomas-Fermi radius $2 R_{z}=180$ $\mu$m \cite{Kasamatsu}. A rotating drive is turned on by giving the value of $\Omega = 0.7 \omega_{\perp}$ and non-adiabatically varying a small anisotropy parameter of the trap as $\epsilon(t) = \Delta \epsilon t \Theta (t_{\rm off} - t)$ with a rate $\Delta \epsilon = 1.25 \times 10^{-3}/$ msec and a switch-off time $t_{\rm off}=20$ msec, where $\Theta(x)$ is a step function. 

\begin{figure}
\includegraphics[height=0.308\textheight]{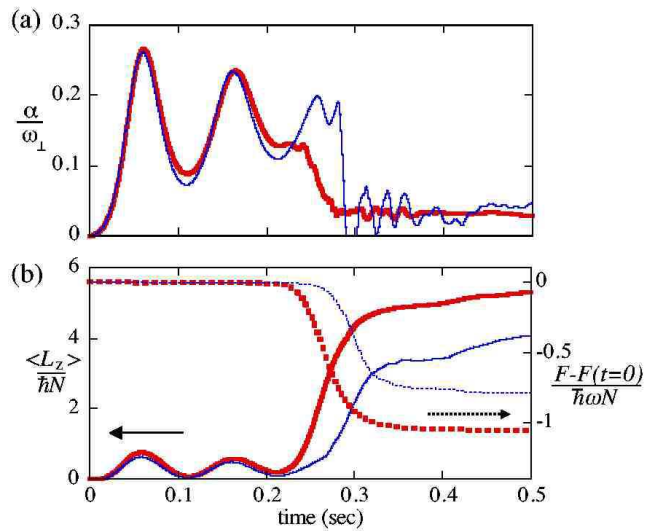}
\caption{(Color online) Time development of parameters in the 2D and 3D simulations. (a) The deformation parameter $\alpha$. In both plots, the 3D results are the thick lines and the 2D results are thin lines. (b) The angular momentum per atom $\langle L_{z} \rangle$ (solid curve) and the total free energy $F$ (dotted curve). The values of the parameters are $\lambda = 9.2^{-1}$, $\Omega = 0.7\omega_{\perp}$, and $u = 54760$ for the 3D simulation and $u_{\rm 2D} = 440$ for the 2D simulation.}
\label{iroiro}
\end{figure}
To gain insight on the overall dynamics of vortex lattice formation, we first show in Fig. \ref{iroiro} the time development of the deformation parameter $\alpha = - \Omega (\langle x^{2} \rangle - \langle y^{2} \rangle)/(\langle x^{2} \rangle + \langle y^{2} \rangle)$, the angular momentum per atom $\langle \hat{L}_{z} \rangle$, and the total free energy $F= \langle \hat{H}-\Omega \hat{L}_{z} \rangle$ with $\hat{H}=-\nabla^{2}+\tilde{V}+u|\Psi|^{2}/2$, where $\langle \hat{A} \rangle = \int d {\bf r} \Psi^{\ast} \hat{A} \Psi$. From Fig. \ref{iroiro}(a), the deformation parameter initially has a damped quadrupole oscillation as observed in the experiment \cite{Madison2}. After two oscillation periods, $\alpha$ falls abruptly to a value just below 0.05 and settles down to a steady small-amplitude oscillation. Concurrently, a vortex lattice forms during the rapid increase in the angular momentum, and the simultaneous decrease in the free energy as seen in Fig. \ref{iroiro}(b). This formation time, which depends on the value of $\gamma$, is in good agreement with the experimentally observed value of 250 msec \cite{future}. 

\begin{figure*}
\includegraphics[height=0.51\textheight]{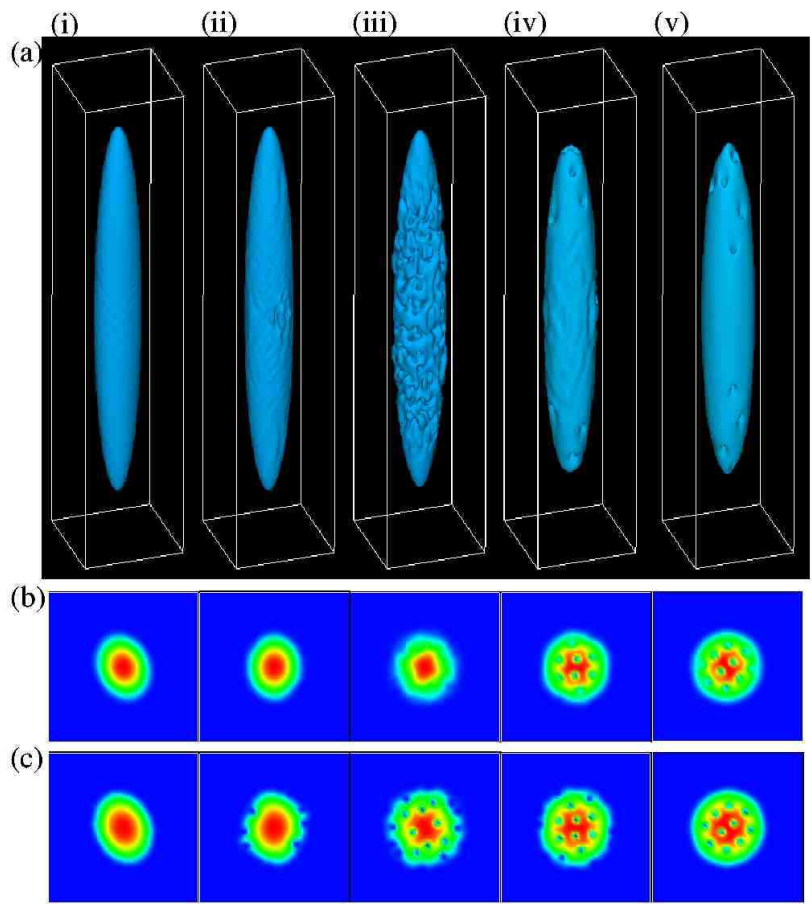}
\caption{(Color online) Time development of vortex formation in 3D. (a) The constant surface density $|\Psi|^{2} = 2 \times 10^{-5}$. (b) The column density $\rho(x,y) = \int dz |\Psi(x,y,z)|^{2}$. (c) The cross-sectional density $|\Psi(x,y,0)|^{2}$. The times are (i) 95 msec, (ii) 220 msec, (iii) 270 msec, (iv) 365 msec, (v) 500 msec. The box dimensions along $x$, $y$, and $z$ axis are -14.5 to +14.5, -14.5 to +14.5, and -72.5 to +72.5, respectively, in units of $b_{\rm ho}=0.728$ $\mu$m ($116 \times 116 \times 580$ discretized space).}
\label{densitydev}
\end{figure*}
More detailed features are revealed in 3D images of the condensate density $|\Psi|^{2}$. The plots in Figure \ref{densitydev}(a) show the time development of a constant surface of the condensate density. Before the abrupt decrease in $\alpha$ in Fig. \ref{iroiro}(a), the condensate undergoes a quadrupole deformation [Fig. \ref{densitydev}(a-i)]. After that, the condensate surface begins to show stripe-like surface waves that propagate along the surface and are nearly parallel to the $z$-axis [Fig. \ref{densitydev}(a-ii)]. The resultant surface waves lead to a state with turbulent density fluctuations as shown in Fig. \ref{densitydev}(a-iii). Most of these density fluctuations rapidly decay because of the dissipation, but some of them evolve into vortex lines that later penetrate into the condensate [Fig. \ref{densitydev}(a-iv)] and form a triangular lattice [Fig. \ref{densitydev}(a-v)]. The surface wave instability occurs initially in the middle of the elongated axis (near $z \simeq 0$). If the 3D condensate is regarded as an assembly of the sliced 2D condensates in the $x$-$y$ plane, the central slice has the largest raduis and the largest 2D coupling constant $u_{\rm 2D}$. Then. if 2D analysis is applied, as in Refs. \cite{Kasamatsu,Isoshima}, the plane with the largest coupling constant would be the first to have an instability, because the corresponding critical rotation frequency is smallest in that plane. This is consistent with the finding that the surface wave instability starts at $z=0$. 

\begin{figure*}
\includegraphics[height=0.45\textheight]{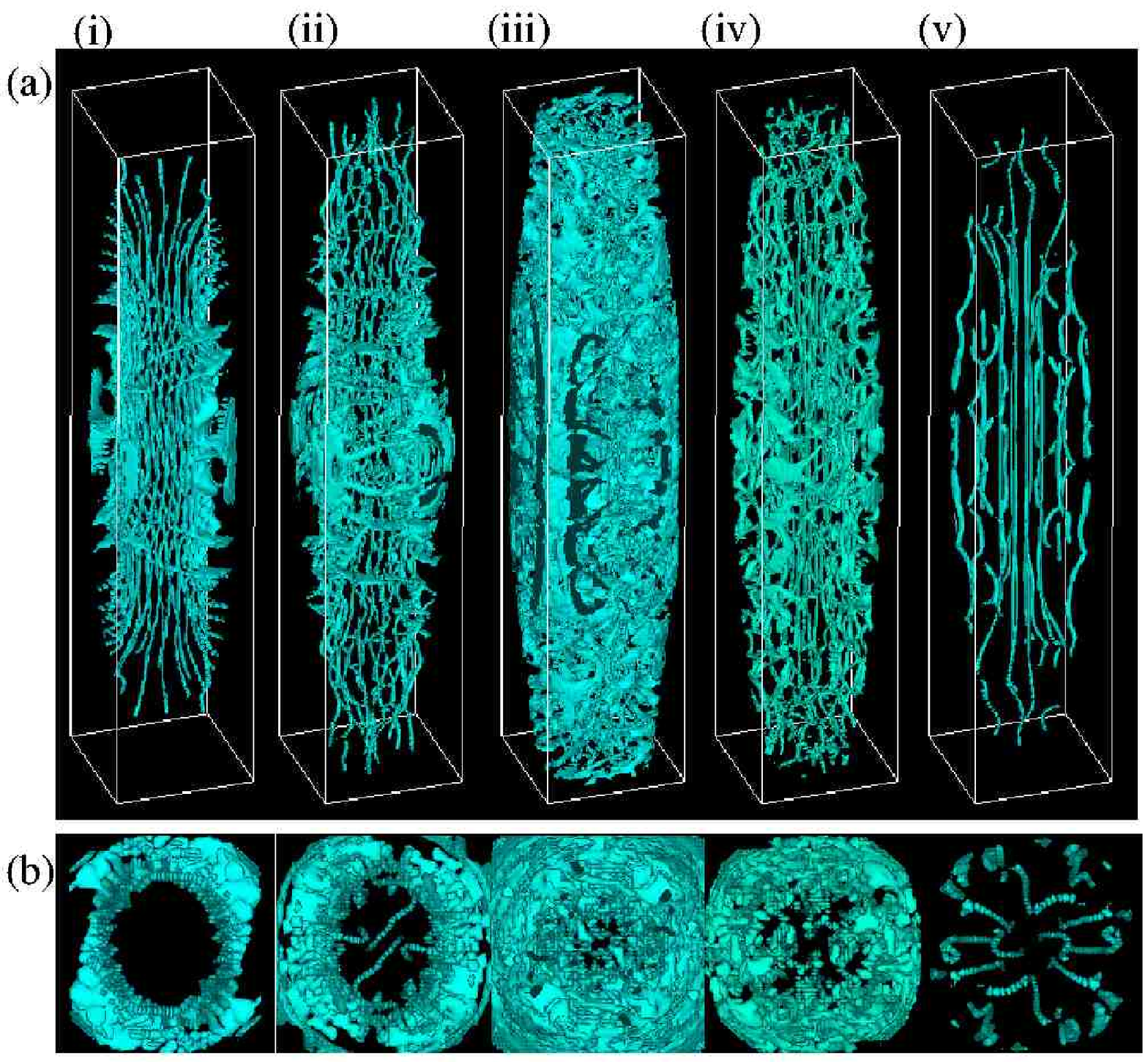}
\caption{(Color online) Time development of a surface of constant superfluid flow velocity with $|{\bf v}|=3.2$. (a) Snapshots in 3D. (b) Top views of (a). The times of (i) to (v) are the same as those in Fig. \ref{densitydev}.}
\label{velodev}
\end{figure*}
The most remarkable feature of the 3D simulations is that the vortex lines are bent. This feature can be seen in their column density profiles $\rho(x,y) = \int dz |\Psi|^{2}$ shown in Fig. \ref{densitydev}(b), which have often used to visualize the vortices in experiments \cite{Madison,Hodby,Madison2}. The column density profiles differ from the cross-sectional density at $z=0$ in Fig. \ref{densitydev}(c); in particular, the density dips characteristic of the vortex cores, the small dark circles in the image, are hard to discern just after the vortex penetration starts as shown in Fig. \ref{densitydev}(b-iii) and (c-iii). This is due to the bending of the vortex lines. After 350 msec, the blurred image of the vortex cores gradually becomes clear [Fig. \ref{densitydev}(b-iv)]. We also found that the locations of the vortex cores in the cross-sectional density had an unsteady motion, a behavior that was not seen in the column density \cite{movie}. This motion indicates the existence of the Kelvin waves propagating along the vortex lines. Near equilibrium, the visibility of the vortices are still different from those in Fig. \ref{densitydev}(b-v) and (c-v), because the vortex lines still bend near the edge of the condensates, which is consistent with the previous equilibrium arguments \cite{Ripoll}. Animation of the above simulation is in Ref. \cite{movie}.

In the previous (2D) study \cite{Tsubota,Kasamatsu}, we investigated the dynamics of the phase field $\theta = {\rm arg}\Psi$. This is difficult in the 3D case; neverthless, the fact that the value of $|{\bf v}|$ diverges at the vortex core allows one to study the evolution of the phase field by calculating the superflow velocity $|{\bf v}| =|\nabla \theta| = |(\Psi^{\ast} \nabla \Psi - \Psi \nabla \Psi^{\ast})/2i |\Psi|^{2} |$. Figure \ref{velodev}(a-i) and (b-i) shows that an array of vortex lines surrounds the condensate and precesses around it even before the surface wave instability occurs. These lines are ghost vortices \cite{Tsubota} that quickly appear near the radial Thomas-Fermi boundary $R_{\perp}(z)$ after the rotation is turned on because their creation energy, which is proportional to the density, is almost negligible. The present 3D simulation reveals in Fig. \ref{velodev}(a-ii) that the ghost vortices vibrate rapidly before they penetrate into the condensate. This is understood by the fact that the precession frequency of an off-centered 2D vortex is proportional to $R_{\perp}^{-2}$ \cite{Fetter}. By treating the 3D condensate as an assembly of 2D slices, one can understand that the precession of a ghost vortex near $R_{\perp}(z)$ at large $|z|$ is faster than that at $z \simeq 0$. Therefore, a density inhomogeneity along the $z$-axis induces a mismatch of the precession frequency at each $z$, which causes the ghost vortex lines to twist and excites Kelvin waves. At the unstable stage, the vibrating ghost vortices give rise to violent density fluctuations in Fig. \ref{densitydev}(a-iii), and the velocity field on the outskirts of the condensate enters the strongly turbulent regime. This feature appears to be unique to cigar-shaped condensates: spherically-symmetric condenesates and pancake-shaped condensates have nearly straight penetrating vortices \cite{Sasa}. After this turbulent stage, the dissipation removes some velocity fluctuations and makes the Kelvin modes damp out. At the quasi-stationary stage in Fig. \ref{velodev}(a-v) and (b-v), the bent vortex lines inside the condensate move slowly and settle down to the lattice structure, while the surrounding ghost vortices precess and eventually disappear outside. 

We briefly compare the 2D and 3D results. In Fig. \ref{iroiro}(a), the quadrupole oscillation frequency in 2D is the same as that in 3D, but the 2D result shows lingering damped oscillations after the surface wave instability. The 3D case is different because a given value of $\alpha$ in the $x$-$y$ plane is integrated over $z$, and thus the oscillation in $\alpha$ should vanish when the condensate density becomes turbulent. Another difference between the 2D and 3D cases is that the 3D system has a larger quasi-steady value of the angular momentum after about 0.3 seconds [Fig. \ref{iroiro}(b)]. The reason for this difference is likely the following. After the vortex penetration, the centrifugal force expands the condensate in the radial direction. In the 3D calculation, the conservation of particle number decreases $R_{z}$ and results in an increase in the effective 2D coupling constant $u_{\rm 2D}$. In contrast, this coupling constant is fixed in the 2D case. The increase in $u_{\rm 2D}$ allows the penetration of more vortices \cite{Kasamatsu}, leading to a slight increase of the final angular momentum. Therefore, except for vibrations of the vortex lines, the overall dynamics observed in the experiments may be captured qualitatively in the 2D simulations. 

In conclusion, we did fully 3D numerical simulations of the GP equation to investigate the dynamics of vortex lattice formation in a rotating BEC. We found that the generalized GP equation with a phenomenological dissipation term can accurately describe the experimental observations reported in Ref. \cite{Madison2}. The 3D simulations revealed that, just before the vortex penetration, the condensate underwent a strongly turbulent process caused by the Kelvin wave excitation of the ghost vortices. These excitations were generated by a density inhomogeneity along the $z$-axis. The vortices were found to bend immediately after the ghost vortices developed into real vortices, which reduced the vortex visibility in the column density profile. 

The first author is grateful to M. Kobayashi for his technical support. K.K. and M.T. acknowledge support by a Grant-in-Aid for Scientific Research(Grant No. 15$\cdot$5955 and No. 15341022, respectively) by the Japan Society for the Promotion of Science.



\begin{thebibliography}{99}
\bibitem{tyuu}
In a separate simulation of the GPE without dissipation, we found that damping of a condensate motion is caused by the stochastic process, in which the energy of low-energy (long-wavelength) excitation modes is transferred irreversibly into high-energy (short-wavelength) modes. We found that this process is strongly affected by the mesh size of the numerical simulations. Also, without dissipation, the vortices appeared near the condensate surface but the formation of a lattice did not occur within the relevant time scale (i.e., a few hundred milliseconds).
\bibitem{future}
The $\gamma$ dependence of the dynamics will be reported elsewhere.
\end{thebibliography}
\end{document}